%% file: CBELG.tex
\documentclass[useAMS,usenatbib]{mn2e}

\usepackage{amsfonts,amsmath,amssymb,mathrsfs,times,graphicx}

\title[Two phase galaxy formation: The Evolutionary Properties of Galaxies]{Two phase galaxy formation: The evolutionary properties of galaxies}

\author[M. Cook et al.]{M. Cook$^{1,2}$\thanks{E-mail:cook@sissa.it (MC)}, E. Barausse$^{5,1}$, C. Evoli$^{1}$, A. Lapi$^{3,1,4}$, G.L. Granato$^{4,2}$\\
$^{1}$Astrophysics Sector, SISSA/ISAS, Via Beirut 2-4, I-34014 Trieste, Italy\\
$^{2}$INAF, Osservatorio Astronomico di Padova, Vicolo dell' Osservatorio 5, I-35122 Padova, Italy\\
$^{3}$Dept. of Physics, Univ. di Roma `Tor Vergata', Via della Ricerca Scientifica 1, I-00133 Rome, Italy\\
$^{4}$INAF, Osservatorio Astronomico di Trieste, Via G.B. Tiepolo 11, I-34131 Trieste, Italy\\
$^{5}$Centre for Fundamental Physics, University of Maryland, College Park, MD 20742-4111, USA}

\newcommand{\lcdm}{$\Lambda$CDM }

\newcommand{\hi}{$HI$ }
\newcommand{\hii}{$H_{2}$ }

\begin{document}

\maketitle

\label{firstpage}

\begin{abstract}

We use our model for the formation and evolution of galaxies within a two-phase galaxy formation scenario, showing that the high-redshift domain typically supports the growth of spheroidal systems, whereas at low redshifts the predominant baryonic growth mechanism is quiescent and may therefore support the growth of a disc structure. Under this framework we investigate the evolving galaxy population by comparing key observations at both low and high-redshifts, finding generally good agreement. By analysing the evolutionary properties of this model, we are able to recreate several features of the evolving galaxy population with redshift, naturally reproducing number counts of massive star-forming galaxies at high redshifts, along with the galaxy scaling relations, star formation rate density and evolution of the stellar mass function. Building upon these encouraging agreements, we make model predictions that can be tested by future observations. In particular, we present the expected evolution to z=2 of the super-massive black hole mass function, and we show that the gas fraction in galaxies should decrease with increasing redshift in a mass, with more and more evolution going to higher and higher masses. Also, the characteristic transition mass from disc to bulge dominated system should decrease with increasing redshift.
\end{abstract}


\input{introduction}

\input{model}

\input{results}

\input{conclusions}

\appendix
\section{Tabulated model predictions}

Within this appendix we tabulate the results for the black hole mass function at $z=2$ and the stellar-to-gas fraction up to $z=5$ where currently there are no strong constraints, but with future studies, these model predictions may be compared with observations.

%%Results from fig.8, right panel
\begin{table*}[h]
\centering
\caption{Values of the model outputs for the black hole mass function at $z=2$ (see Fig.8).}
\begin{tabular}{lcccc}
\hline
\hline
$\log(M_{bh}) [M_{\odot}]$ & $\phi(M_{bh}) ~[Mpc^{-1}]$ & $\phi(M_{bh})_{max} ~[Mpc^{-1}]$ & $\phi(M_{bh})_{min} ~[Mpc^{-1}]$\\
\hline
 $7.15$ & $-2.28$ & $-2.52$ & $-2.10$ \\
 $7.48$ & $-2.15$ & $-2.33$ & $-1.93$ \\
 $7.81$ & $-2.20$ & $-2.40$ & $-1.99$ \\
 $8.14$ & $-2.41$ & $-2.59$ & $-2.20$ \\
 $8.47$ & $-2.64$ & $-2.85$ & $-2.43$ \\
 $8.80$ & $-2.90$ & $-3.11$ & $-2.69$ \\
 $9.13$ & $-3.29$ & $-3.50$ & $-3.08$ \\
 $9.46$ & $-3.92$ & $-4.12$ & $-3.73$ \\
 $9.79$ & $-4.91$ & $-5.07$ & $-4.70$
\\
\hline
\end{tabular}
\end{table*}

%%Results from fig.8, right panel
\begin{table*}[h]
\centering
\caption{Values of the model outputs for the stellar-to-gas fractions (see Fig.9).}
\begin{tabular}{ccccccc}
\hline
\hline
$\log(M_{s}) [M_{\odot}]$  & $\log(M_{g}/M_{s})_{z=0}$ & $\log(M_{g}/M_{s})_{z=1}$ &  $\log(M_{g}/M_{s})_{z=2}$ & $\log(M_{g}/M_{s})_{z=3}$ & $\log(M_{g}/M_{s})_{z=4}$ & $\log(M_{g}/M_{s})_{z=5}$\\
\hline
$7.20$ & $1.20$ & $1.18$ & $1.03$ & $0.94$ & $0.84$  & $-0.29$ \\
$7.65$ & $0.93$ & $0.88$ & $0.74$ & $0.72$ & $-0.07$  & $-0.38$ \\
$8.10$ & $0.69$ & $0.62$ & $0.58$ & $0.50$ & $-0.87$  & $-1.24$ \\
$8.55$ & $0.51$ & $0.39$ & $0.36$ & $-1.02$ & $-1.27$  & $-1.25$ \\
$9.00$ & $0.25$ & $0.14$ & $-1.12$ & $-1.63$ & $-1.36$  & $-1.24$ \\
$9.45$ & $0.01$ & $-0.18$ & $-1.88$ & $-1.50$ & $-1.39$  & $-1.17$ \\
$9.90$ & $-0.24$ & $-0.47$ & $-1.90$ & $-1.53$ & $-1.41$  & $-1.18$ \\
$10.35$ & $-0.46$ & $-1.51$ & $-1.82$ & $-1.53$ & $-1.30$  & $-1.17$ \\
$10.80$ & $-0.61$ & $-2.20$ & $-1.83$ & $-1.90$ & $-1.26$ & $-0.83$
\\
\hline
\end{tabular}
\end{table*}

\section*{Acknowledgments}

We thank F. Shankar and A. Schurer for stimulating discussions which helped the progress of this work. MC thanks L. Paulatto for considerable computational assistance. MC has been supported through a Marie Curie studentship for the Sixth Framework Research and Training Network MAGPOP, contract number MRTN-CT-2004-503929. EB acknowledges support from NSF Grant No. PHY-0603762. We also thank the annonymous referee for useful comments which helped the clarity of this work.

\bsp

\label{lastpage}

\end{document}

%% file: introduction.tex
\section{Introduction}

The current paradigm of cosmological structure evolution is outlined by the \lcdm model: Providing a remarkably successful framework for interpreting a wealth of observations of cosmic structure evolution over the majority of the duration of the Universe. This model is capable of reproducing the cosmic microwave background radiation fluctuations (Spergel et al. 2007), the large scale clustering of galaxies (Eisenstein et al. 2005, \& references therein), the cosmic shear field measured through weak gravitational lensing (Hoekstra et al. 2006 \& references therein), small scale power spectrum of Lyman-alpha forest sources (Jena et al. 2005), the properties of galaxy clusters (Allen et al. 2004 \& references therein) along with several other key observations of large scale cosmological structures. However, despite these merits, on galaxy scales the assembly of baryonic material within virialised dark matter (DM) haloes has had more mixed successes: Due to the complex processes, often non-linear and dissipative, which operate on scales well below the resolution of the model ('sub-grid' physics), in order to model the evolution of baryonic material within DM haloes one is required to adopt analytic prescriptions and make several important assumptions concerning the geometry of the forming system (see Zavala, Okamoto \& Frenk, 2008).

Early endeavors to model the cosmological evolution of luminous structures came from White \& Rees (1978) and Blumenthal et al. (1984), whereby galaxies form when gas cools and condenses within the centres of hierarchically evolving DM haloes. Attempts to model and interpret the evolutionary properties of galaxies within the first generations of semi analytical models (SAMs) showed promising qualitative agreements to observations (Kauffmann et al. 1993, 1998, Cole et al. 1994, 2000, Somerville \& Primack, 1999). However, in the past decade it has become clear beyond reasonable doubt that significant tensions between SAM predictions and fundamental observations exist, most notably in three major areas: Firstly the issue of 'overcooling' ('quenching'), which has several manifestations, large DM haloes are observed to be low in baryonic mass and contain typicaly 'red and dead' early-type galaxies, resulting in a sharp cutoff in the high-mass end of the stellar mass function, unlike that for the DM haloes (Bell et al. 2003a, Benson et al. 2003, see Somerville et al. 2008b for a discussion). Secondly, 'downsizing', or 'anti-hierarchical' evolution of baryonic structures (Cowie et al. 1996), whereby massive star forming systems and associated SMBHs shined mostly at high redshifts, while smaller objects show longer lasting activity (see also Fontanot et al. 2009 for details) which appears contrary to naive expectations for the 'bottom up' growth of DM structure. Finally, the 'dwarf galaxy', or 'substructure' problem, whereby the number of low mass galaxies predicted by models is significantly more than is observed (see Mo et al. 2005, Moore et al. 1999).

Theoretical attempts to interpret these somewhat puzzling properties of galaxies motivated a second generation of SAMs, which evoked strong feedback from a central supermassive black hole (SMBH) in order to quench star formation at late times by suppression of cooling, predominantly in the larger galaxies (see Croton et al. 2006, Bower et al. 2006, Baugh et al. 2006, see also Granato et al. 2004), generating a marked improvement over previous incarnations, but several tensions remained (see Monaco et al. 2007). The current state-of-the-art SAMs include the energetic effects of growing central SMBH, the effects of hot and cold accretion (Dekel \& Birnboim, 2006, Cattaneo et al. 2006, Somerville et al. 2008b) or a flat stellar initial mass function (IMF) during starburst activity (Baugh et al. 2005), the suppression of cooling and collapse due to an ionizing UV background (see Gnedin, 2000, Somerville 2002, Benson et al. 2002), thus steadily increasing the degrees of freedom in order to improve agreement with observational constraints. Progress is currently being made in developing SAMs with added layers of physical descriptions, including spatially resolved modeling (see Stringer \& Benson, 2007, Dutton \& van den Bosch, 2009, Cook et al. 2009a (hereafter C09a)), multi-phase ISM physics (Dutton \& van den Bosch 2009, Cook et al. 2009b (hereafter C09b)) in order to increase predictability of models without significantly increasing their number of free parameters.

\begin{figure}
  \centering
  \includegraphics[type=eps,ext=.eps,read=.eps, height=8.5cm, angle=90]{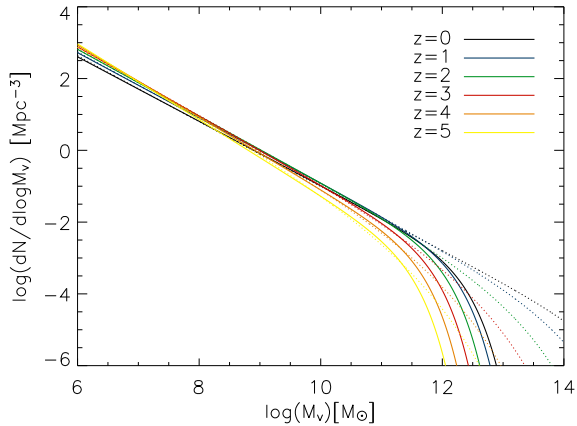}\\
  \caption{The evolution of the galaxy halo mass function (GHMF) shown with the solid curves, from $0<z<5$ and the Sheth-Tormen mass function (STMF) shown with the dotted curves. As can be seen, in all redshift ranges the GHMF resembles the STMF until approximately $10^{12}M_{\odot}$ where the probability for multiple galaxy halo occupation grows and thus the single galaxy mass probability diminishes, resulting in the exponential cutoff.}
\end{figure}

Since the first generation of SAMs were developed, observational studies have undergone many revolutions due to increased sensitivity, increased wavelength coverage, and automated survey methods. Many of the observational constraints coming as a surprise to the community: At low redshifts, detailed constraints on the stellar mass function (Cole et al. 2001, Bell et al. 2003a) improved model parameter refinements, however, analysis of high redshift star-forming galaxies ($z \ge 2$) opened a window to study the properties of galaxies when the Universe was under 20\% it's current age, lyman break galaxies (Steidel et al. 1996), Sub-mm galaxies (Smail et al. 1997) and Ly-$\alpha$ galaxies (Hu, Cowie \& McMahon, 1998) were generally interpreted as being dusty starbursting systems, with detailed analysis showing that the star formation rate density of the Universe at $z>2$ remains flat (in contrast with the original determination of Madau et al. 1996).
Also, measurements of the mass distribution of high-z galaxies revealed a substantial population of extremely massive galaxies at $z > 1$ (Cimatti et al. 2002, Drory et al. 2003, Kodama et al. 2004, Bundy et al. 2005) in sharp contrast to the original hierarchical picture of structure growth. Current observations of stellar mass now extend to $z \approx 5$ (Drory et al. 2005, Fontana et al. 2006, Elsner et al. 2008, Per�z-Gonz�lez et al. 2008, Marchesini et al. 2009), and theoretical models must attempt to interpret these results physically whilst simultaneously making predictions about the black hole growth (Hopkins et al. 2006) for which observations are complete to high redshifts, and the scaling relations of galaxies and SMBHs (see Woo et al. 2008) along with the 'archeological' constraints on the evolutionary properties of galaxies (see Gallazzi et al. 2005). It has been shown that theoretical models have had mixed successes, with no model currently able to consistently predict all observations (see Kitzbichler \& White, 2007, Marchesini \& van Dokkum, 2007, Somerville et al. 2008b, Fontanot et al. 2009).

\begin{figure*}
\begin{center}
  \includegraphics[type=eps,ext=.eps,read=.eps, height=18.0cm, angle=90]{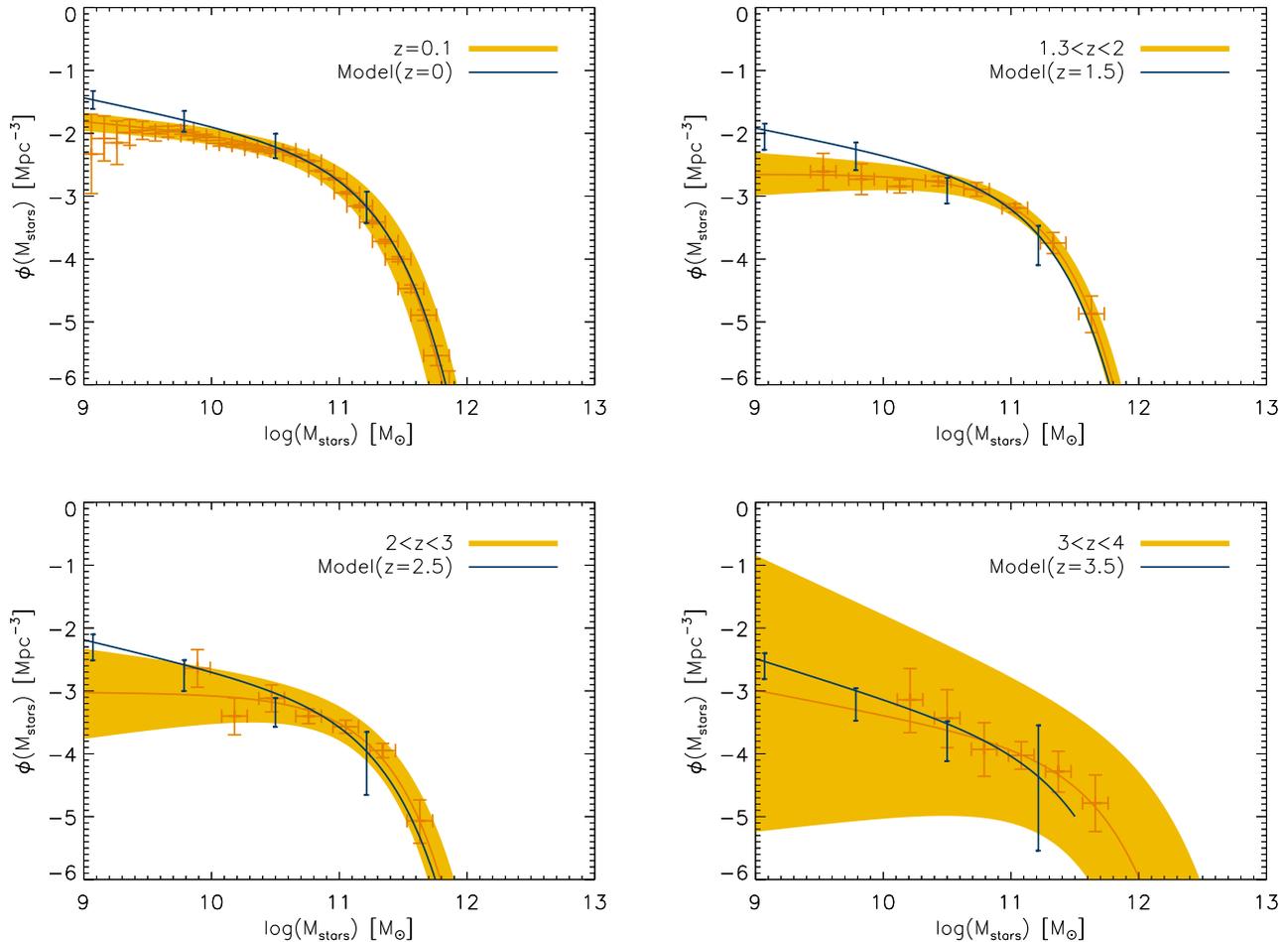}\\
  \caption{The stellar mass function evolution: Locally (top left) plotted against the determinations of Cole et al. 2001 showing good agreement with both the cutoff and normalisation. At $z>0$ compared to Marchesini et al. 2009, showing the $z\approx1.5$ relation (top right), reproducing the high mass cutoff and knee, but slightly overproducing the number of low-mass galaxies, the $z\approx2.5$ relation (bottom left), showing an overall good fit within the observational range, and the $z\approx3.5$ relation (bottom right), showing good agreement within the data ranges, but not showing a good approximation to a Schechter function fit. Yellow shaded regions represent the errors in the functional fits (orange lines), and the blue error bars in the model outputs representing the Poisson uncertainty in the mean averages. We note that slight discrepancies in the low mass end may be somewhat attributed to observational biases, and environmental effects which we neglect to model (see \S4 for a further discussion.)}
\end{center}
\end{figure*}

Despite several differences in the detailed 'sub-grid' recipes adopted by different groups, current SAMs all follow the same general framework as originally proposed by White \& Rees, 1978 and adopt the same original assumptions: (i) Gas cooling and condensation with in DM haloes, at any epoch, results in the dissipationless formation of a self-gravitating gaseous disc which undergoes mild star formation. (ii) The main driver for starburst activity is the merging of these gas-rich discs (wet mergers) which also provides the main channel for the formation of spheroidal structures (Cole et al. 1991). The resultant 'disc merger' framework provides the basis for most current SAMs (see Somerville et al. 2008b for a review). However, in our view these strict assumptions may be the underlying cause of several tensions between models and observations (see Mo \& Mao, 2004, C09a), notably the tendency for baryonic material to follow the hierarchical evolution of DM haloes, the difficulty in producing massive galaxies at early times which later passively evolve, and archeological issues relating to the structure of DM haloes and the observed baryon fraction in galaxies (see the aforementioned references).

Within this work, we develop the model outlined in C09a, C09b, where, motivated by the above-mentioned tensions between theoretical models and observations we proposed a model which differs substantially from the standard 'disc merger' framework: We envisage that the fundamental dichotomy between galactic spheroid and disc components is a manifestation of two distinct modes of the evolution of baryonic matter, ultimately driven by the two-phase structural evolution of DM haloes (see Zhao et al. 2003a, Mo \& Mao, 2004, Diemand et al. 2007, C09a, C09b): An early '\emph{fast collapse}' phase, where the DM core structure is constructed through a series of violent merger events, corresponding to an epoch where baryonic material effectively dissipates angular momentum upon collapse to directly form a spheroid-SMBH system, and a late '\emph{slow collapse}' phase, where potentially large amounts of material are added to the halo outskirts little affecting the central regions, giving rise to the quiescent growth of disc structures around the pre-formed spheroids.

Dark matter mergertrees outline the merging rates of DM haloes and are well constrained by simulations. However, ultra high-resolution simulations are required in order to analyse the structure and substructure evolution within the merging DM haloes. Thus, until recently, oversimplified analytical recipes are commonly used (Chandraseakhar, 1943), not accounting for several important effects. Recently, increased numerical resolution and substructure analysis has allowed for some advances in determining the evolution of subhaloes after they have entered a parent halo (which is of upmost importance for baryonic physics), showing that in general, the evolution of the structure of a galaxy-sized DM halo evolves in two-phases.

More specifically, analysis of the cosmological evolution of virialised structures is long standing, from observational clustering studies, through detailed cosmological simulations (Springel et al. 2005), and monte-carlo algorithms tuned to reproduce these results (see Parkinson et al. 2008 \& references therein), however, until relatively recently, determining the detailed evolution of substructure within DM haloes \emph{after} they merge has been somewhat overlooked. Recent increases in numerical resolution within N-body simulations have begun to analyse the detailed structural evolution of haloes within cosmological volumes (Zhao et al. 2003a, 2003b, Diemand, Kulhen \& Madau, 2007, Hoffmann et al. 2007, Ascasibar \& Gottloeber, 2008), showing that two distinct phases of structural evolution are found, an early 'fast collapse' phase, followed by a late 'slow collapse' phase. This has also prompted several works to show how typical double power-law DM halo density profiles may be generated (see Lu et al. 2006, Lapi \& Cavaliere, 2009). This theoretical idea has also been hinted upon in the Millennium Galaxy Catalogue bulge-disc decomposition analysis of Driver et al. 2006.

Motivated by these issues, within this contribution we expand the model presented in C09a and C09b, which comprises a natural extention to the spheroid-SMBH co-evolution model presented in Granato et al. 2004, (see also Granato et al. 2001, Lapi et al. 2006, 2008, Mao et al. 2007), and focus on several 'problem plots' for current SAMs under the disc-merger framework. We essentially inherit the results of these papers here. By self-consistently outputting galaxy properties at several different redshits for a representative sample of galaxies generated by our model, we are able to model the evolutionary development of baryonic material within our models, comparing the fundamental relations in order to constrain the key physical mechanisms governing galaxy formation. Focusing on the evolution of the galaxy scaling relations for discs and spheroids, the evolution of the mass functions for both SMBHs and galaxies, the cosmological star formation rate density, the cosmological evolution of the most massive galaxies and the archeological stellar populations of local galaxies.

The plan of this paper is as follows; in \S2 we overview the physical model, highlighting important points and modifications to previous works, in \S3 we describe the methods in order to extract observable quantities and present the results for the evolving galaxy population, we conclude and summarise our findings in \S4, highlighting the successes and limitations of our approach. Throughout the paper we adopt the standard $\Lambda$CDM concordance cosmology, as constrained by \textsl{WMAP} 5-year data (Spergel et al. 2007). Specifically, we adopt a flat cosmology with density parameters $\Omega_M=0.27$ and $\Omega_{\Lambda}=0.73$, and a Hubble constant $H_0=70$ km s$^{-1}$ Mpc$^{-1}$.

\begin{figure}
  \centering
  \includegraphics[type=eps,ext=.eps,read=.eps, height=8.5cm, angle=90]{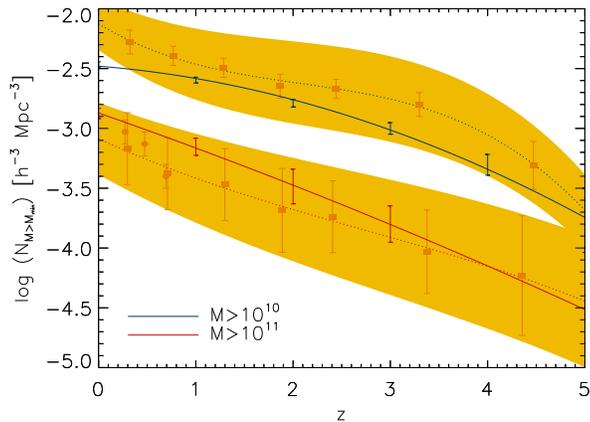}\\
  \caption{The number density of galaxies with stellar masses $M>10^{10}M_{\odot}$ and $M>10^{11}M_{\odot}$ produced by Drory et al. 2005 and compared with model prescriptions. Yellow shaded regions represent the error in the functional fits (orange lines), and the blue error bars in the model outputs representing the Poisson uncertainty in the mean averages. Due to the large star-burst activity at high-z we are able to construct massive galaxies readily and produce good agreements to the data.}
\end{figure}

%% file: model.tex
\section[]{Overview of the Model}

\begin{table*}
\centering
\caption{Values of the free parameters of our model.}
\begin{tabular}{lcccc}
\hline
\hline
Description &  Symbol & Fiducial value & Impact on this work\\
\hline
%BH radiative efficiency         &   $\eta$                   &  $0.15$ &  & Mild\\
SN feedback efficiency (bulge)         &   $\epsilon_{\rm SN,b}$   &  $0.5$  & Strong\\
SN feedback efficiency (disc)        &   $\epsilon_{\rm SN,d}$   &  $0.8$  & Strong\\
Reservoir growth rate           &   $A_{res}$               & $10^{-3} M_{\odot}{\rm yr}^{-1}$    & Strong\\
QSO feedback efficiency         &    $f_{\rm h,QSO}$                &  $10^{-4}$             & Strong\\
Radio feedback efficiency         &    $f_{\rm h,radio}$                &  $1$             & Strong\\
%Gas velocity dispersion  &   $\sigma_{\rm gas}$     &  $6$ km s$^{-1}$        & Eq.~31 & Weak\\
%Normalization constant   &   $Q$                    &  $1.5$                  & Eq.~31 & Weak\\
%Disc stability parameter        &       $\alpha_{crit}$     &  $0.8$                & Eq.~38      & Weak\\
Viscous accretion rate & $k_{acc}$ & $10^{-2}$  & Weak\\
Radio mode accretion rate & $k_{\rm radio}$ & $6\times10^{-6} M_{\odot}{\rm yr}^{-1}$  & Weak
\\
\hline
\end{tabular}
\end{table*}

We refer the reader to C09a and C09b for a detailed description of the model details. However, in order to preserve clarity we review the main model features here.

For the dark matter merging and accretion evolution, we adopt an extended Press-Schechter formalism based on the binary mergertree of Cole et al. 2000, as modified by Parkinson et al. 2008. This algorithm has been shown to reproduce halo merging and accretion statistics obtained from cosmological numerical simulations (Springel et al. 2005). We use these mergertrees by extracting the main-progenitor mass accretion history (MAH) beginning at $z=0$ and moving to progressively higher redshifts in the mergertree, taking the largest progenitor branch at each merger event.

It has been shown in high-resolution simulations, and investigated in analytical dark matter studies, that the structure of DM haloes evolves in two phases, a 'fast' accretion phase at high $z$ and a 'slow' accretion phase at low $z$. These two phase are reflected in the redshift evolution of the concentration parameter $c(z)$, which characterises the halo structure. In our work, we calculate $c(z)$ by means of recent simulation results for the $z=0$ mass-concentration relation (Maccio et al. 2007), coupled to the evolutionary evolution reported in Zhao et al. 2003a,
\begin{equation}\label{11}
  { [\ln(1+c)-c/(1+c)]c^{-3\alpha}  }   \propto {H(z)}^{2\alpha} {M_{\rm vir}(z)}^{1-\alpha}\,,
\end{equation}
where $\alpha$ is a piecewise function (Zhao et al. 2003a). The knowledge of the evolution of $c(z)$ then allows us to distinguish the slow and fast DM accretion phases, which we associate with
two growth mechanisms for the baryonic sector: the fast DM accretion phase giving rise to the the formation
of bulges,  and the slow DM accretion phase giving rise to the the formation
of discs  (Mo \& Mao, 2004, C09).

\begin{figure}
  \centering
  \includegraphics[type=eps,ext=.eps,read=.eps, height=8.5cm, angle=90]{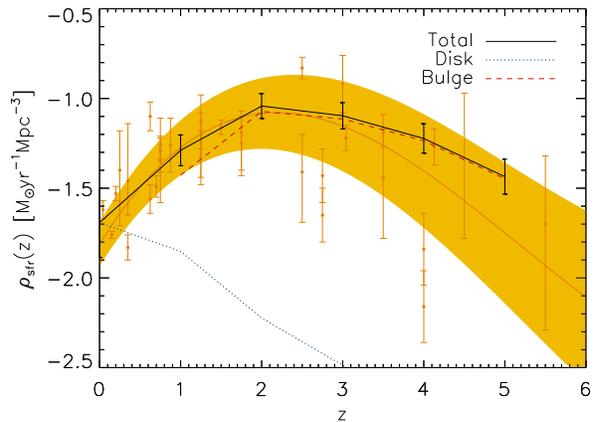}\\
  \caption{The cosmological star formation rate density evolution. Model comparisons with the Somerville et al. 2001 compilation, showing good agreements to observations between $0<z<5$, also plotted are the contributions from the spheroid (red) and disc (blue) components. The yellow shaded region represents the errors in the functional fits (orange line), and the blue error bars in the model outputs representing the uncertainty in the mean averages. These results show, in broad agreement with observations, that elliptical galaxies (and galaxy bulges) form early, and discs form late, with a typical transition at $z\approx1$.}
\end{figure}

More specifically, in order to model the baryonic evolution, we start at a redshift at which the virial mass (given by the MAH) reaches the so-called cooling mass, \textit{i.e.} the virial mass at which  $T_{\rm vir}=10^{4}K$. In fact, below this temperature the $Ly\alpha$ cooling becomes inefficient and baryonic structures cannot form. Following this, over each redshift increment, we allow hot gas to accrete onto the DM halo with rate $\dot{M}_{\rm inf}= f_{\rm coll} \dot{M}_{\rm vir}$, where $f_{\rm coll}$ is the baryonic collapse fraction in the presence of an ionizing UV background (Gnedin et al., 2004, Somerville et al., 2008b)
\begin{equation}\label{1}
   f_{\rm coll}(M_{\rm vir},z) = \frac{\Omega_{b}/\Omega_{m}}{(1+0.26M_{f}(z)/M_{\rm vir})^{3}}
\end{equation}
($M_{f}(z)$ is the filtering mass at a given redshift: see Kravtsov, Gnedin \& Kyplin, 2004, Appendix B).
Also, we include the effects of cold accretion flows, shown to be the predominant mechanism leading to the formation of low-mass systems. Below a critical mass
\begin{equation}\label{1}
   M_{c} = M_{s} \max[1, 10^{1.3(z-z_{c})}]\,,
\end{equation}
where $M_{s}=2\times10^{12} M_{\odot}$ and $z_{c}=3.2$, we assume that all gas accreted onto DM halos is not shock heated to the virial temperature of the DM halo, but streams in on a dynamical time (see Dekel \& Birnboim, 2006, Dekel et al., 2009, Cattaneo et al, 2006).
Thus, in halos below this mass the collapse happens on the dynamical timescale of the system ($t_{\rm coll} = t_{\rm dyn}$), whereas in halos above this mass $t_{\rm coll} = \max[t_{\rm dyn}, t_{\rm cool}]$, where the cooling timescale $t_{\rm cool}$ is computed in a standard way, assuming material is shock heated to the virial temperature.
The effects of this cold accretion is to enhance star formation at high redshifts relative to the scenario where all material is shock heated.

In order to model the baryonic evolution, we suppose that the hot gas phase collapse gives rise, as we have already mentioned, to bulges during the fast accretion phase and to discs during the slow accretion phase:
\[
\begin{array}{ll}
    M_{\rm coll} \rightarrow   M_{\rm b, gas}     & \quad \mbox{[ $z > z_{t}$]} \\
   M_{\rm coll} \rightarrow   M_{\rm d, gas}    & \quad \mbox{ [ $z < z_{t}$]}
\end{array}\]
$z_t$ being the transition redshift between the slow and fast accretion phases.
Thus, we naturally output the growth of a spheroid structure followed by the growth of a disc structure around the pre-formed spheroids.

For $z>z_t$, as gas collapses into the bulge, bursts of star formation occur which force, by radiation drag, part of the cold gas onto a circumnuclear reservoir with low angular momentum, at a rate
\begin{equation}
\dot{M}_{\rm res}= A_{\rm res}\psi_{b}(t)\,.
\end{equation}
The cold gas in this reservoir then becomes eligible to feed a central seed supermassive black hole at an accretion rate
\begin{equation}
    \dot{M}_{\rm bh,QSO} = \min[ \dot{M}_{\rm visc}, \dot{M}_{\rm edd}  ]\,,
\end{equation}
where $\dot{M}_{edd}$ is the Eddington rate and the viscous accretion rate is parameterized as
\begin{equation}
    \dot{M}_{\rm visc} = k_{\rm acc} \frac{\sigma^{3}}{G} \left( \frac{M_{\rm res}}{M_{\rm bh}} \right)\,,
\end{equation}
where $k_{\rm acc}\approx10^{-2}$ is a free parameter with little impact on our results.
Feedback on the growth of baryonic structure comes from two processes. On the one hand,
supernova (SN) explosions
transfer significant energy into the cold ISM, causing it to be re-heated and ejected from the system. Therefore, by considering energy balance in the ISM, we assume that supernova feedback is able to remove gas from the bulge with efficiency
$\epsilon_{\rm SN,b}$ (ranging from 0 to 1, with $\epsilon_{\rm SN,b}=1$ meaning that all of the SN explosion energy is adsorbed by the ISM). This mechanism is most effective in the low mass systems, which presents shallow potential wells from which the ISM can easily escape due to SN explosions.

%Evolution of the TFR - no evolution plot, slope, scatter etc. - with and without adiabatic contraction
\begin{figure*}
  \centering
  \includegraphics[type=eps,ext=.eps,read=.eps, height=18cm, angle=90]{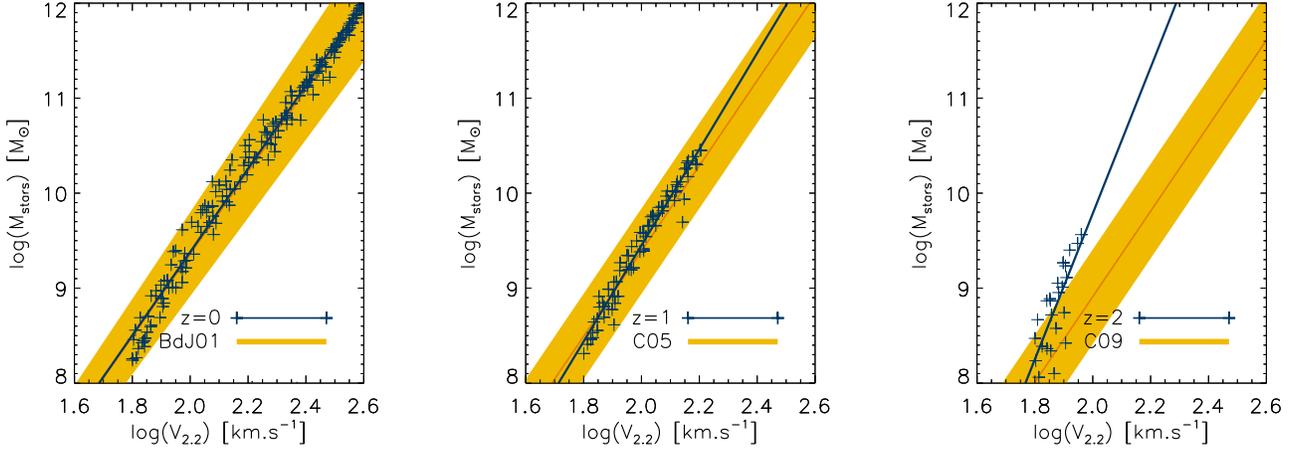}\\
  \caption{The evolution of the stellar mass Tully-Fisher relation, showing model comparisons with Bell \& de Jong, 2001 for local galaxies (left panel) and finding excellent agreement across the observational range, with Conselice et al. 2005 (center panel), finding again good agreement and little evolution from the $z=0$ relation, and finally a tentative comparison with the $z=2$ results of Cresci et al. 2009 (right panel), finding an offset, see \S3.4 for details. Yellow shaded regions represent the scatter in the relation, containing the 95th percentiles, and the mean values are given by the orange lines.}
\end{figure*}

On the other hand, the QSO activity of the central SMBH ejects hot gas and bulge cold gas from the system with an efficiency $f_{\rm h,QSO}$ ($f_{\rm h,QSO}$ ranging from 0 to 1):
\begin{eqnarray}
            \dot{M}_{\rm b,gas}^{QSO}&=&f_{\rm h,QSO}\frac{2}{3} \frac{L_{h}}{\sigma^{2}} \frac{M_{\rm b,gas}}{M_{\rm hot} + M_{\rm b,gas}}\,,\\
            \dot{M}_{\rm hot}^{QSO}&=&f_{\rm h,QSO}\frac{2}{3} \frac{L_{h}}{\sigma^{2}} \frac{M_{\rm hot}}{M_{\rm hot} + M_{\rm b,gas}}\,,
\end{eqnarray}
where $\sigma=\,0.65 V_{\rm vir}$ ($V_{\rm vir}$ being the halo virial velocity), while ${M}_{\rm b,gas}$ and $M_{\rm hot}$
are the masses of the gaseous bulge and of the hot gas phase. This effect is most effective in the large mass systems, where QSO activity is strong.

In addition to the QSO accretion channel, we assume, following Croton et al. 2006, that the SMBH accretes mass also through a quiescent ``radio-mode'' at a rate
\begin{equation}
\dot{M}_{\rm bh\, radio}= k_{\rm radio}\left(\frac{M_{\rm bh}}{10^8 M_\odot}\right) \left(\frac{f_{\rm hot}}{0.1}\right)\left(\frac{V_{\rm vir}}{200 {\rm km/s}}\right)^3\,,
\end{equation}
where $f_{\rm hot}$ is the halo mass in the form of hot gas and $k_{\rm radio}=6\times 10^{-6} M_\odot {\rm yr}^{-1}$ as in Croton et al. 2006. Because of the small value of $k_{\rm radio}$, this mode does not contribute significantly to the SMBH mass evolution. However, following Croton et al. 2006, we assume that the
efficiency $f_{\rm h,radio}$ with which the energy emitted by the SMBH in this mode is adsorbed by the hot gas phase is exactly 1 (\textit{i.e.,} all of the radio mode emission is adsorbed by the hot gas phase).

As the DM halo enters into the relatively quiescent 'slow' accretion phase at $z<z_t$, the DM halo core potential becomes stabilised and we suppose that conditions become sufficient to support the growth of a disc through dissipationless collapse. Thus, gas entering into the DM halo conserves angular momentum and joins a gaseous disc, for which we assume an exponential surface density profile with scale radius is calculated following (Mo, Mao \& White, 1998, equation 29, and C09b equation 31). We adopt for simplicity the same $r_d$ for both the gas and stellar discs, but we have tested that a somewhat larger scale-length for the gas $r_{d,gas}=1.5r_{r,stars}$ (e.g. Somerville et al 2008), does not yield any significant difference in the results discussed here. Star formation in these gaseous discs is expected to take place in molecular clouds (see C09b, section 3.3 for more details on the star formation law that we use) and gives rise to a stellar disc, for which we assume an exponential surface density profile with the same scale radius as the gaseous disc.

It is known that when discs become self-gravitating they are likely to develop bar instabilities, get disrupted and transfer material to the spheroidal component (Christodoulou, Shlosman \& Tohline, 1995). We therefore assume that a stellar or gaseous disk is stable if
\begin{equation}\label{stability}
\frac{V_{\rm c}(2.2 r_d)}{(GM^*_{\rm disc}/r_d)^{1/2}} > \alpha_{\rm crit}^{*}\,\quad *={\rm stars\,, gas}\,,
\end{equation}
where $\alpha^{\rm stars}_{\rm crit}=1.1$ and $\alpha^{\rm gas}_{\rm crit}=0.9$ [see (Mo, Mao \& White, 1998) and references therein]. If we find that discs become unstable, we assume they get disrupted in a dynamical time and transfer their material (either stars or gas) to the bulge components.

Feedback on the disc growth comes again in two fashions. On the one hand, analogously to the bulge case, we assume that
supernova explosions can remove gas from the disc with efficiency $\epsilon_{SN,d}$ (ranging from 0 to 1).
Again, this mechanism is only efficient for small systems.  On the other hand, QSO activity is not generally present in the slow accretion phase, unless the gaseous disc fragments due to bar instability into a spheroidal gaseous component, which immediately forms stars, feeding the reservoir and, through it, the SMBH. However, the radio mode feedback is still present and removes hot gas from the system, thus quenching the collapse of the hot gas into the disc cold gas and indirectly suppressing disc star formation\footnote{In our model we assume the QSO and radio mode feedback do not remove cold gas from the disc, due to the small geometric cross section of the disc relative to the SMBH emission.}

%
%
%  for which we assume an exponential surface density profile for both gas and stars, and to have a scale-radius given by dissipationless collapse. Where $r_{d}(z)$ evolves according to the scaling relation $r_{d}(z) = (2\pi)^{-1/2} (j_{d}/m_{d}) \lambda  r_{v}(z)f(c)^{-1/2} f_{r}(\lambda, c, m_{d} j_{d})$, and for a NFW halo the function $f(\lambda, c, m_{d} j_{d})$ may be exactly determined through (Mo, Mao \& White, 1998).

%
% \begin{equation}\label{4}
%     f(\lambda, c, m_{d}, j_{d}) = 2 \left[ \int_{0}^{ \infty } e^{-u}u^{2}\frac{V_{c}(r_{d}u)}{V_{200}}du \right]^{-1}
% \end{equation}

Finally, in order to account for adiabatic halo response, we take the standard prescription of Blumenthal (1986). In particular, denoting by $M_X(r)$ the mass of a given component '$X$' ($X={\rm b}$ for the bulge, $X={\rm d}$ for the disk, $X={\rm DM}$ for the dark matter and $X={\rm res}$ for the reservoir) enclosed by a radius $r$, from the angular momentum conservation one obtains
\begin{equation}\label{collapse_halo1}
   M_{i}(r_{i})r_{i} = M_{f}(r_{f})r_{f}\,,
\end{equation}
where $r_i$ and $r_f$ are respectively the initial and final radius of the shell under consideration, the initial mass distribution $M_{i}(r_{i})$ is simply given by the NFW density profile, while $M_{f}(r_{f})$ is the final mass distribution. Also, mass conservation easily gives
\begin{eqnarray}
  && M_{f}(r_{f}) = M_{\rm d}(r_{f})+M_{\rm b}(r_f)+M_{\rm DM}(r_{f}) +M_{\rm res}(r_{f})=\nonumber \\
  &&  M_{\rm d}(r_{f})+M_{\rm b}(r_f) +M_{\rm res}(r_{f}) + (1-f_{gal})M_{i}(r_{i})\,,\label{collapse_halo2}
\end{eqnarray}
where $f_{gal}=M_{gal}/M_{\rm vir}$ (with $M_{gal}=M_{\rm d}+M_{\rm b}+M_{\rm res}$). By assuming spherical collapse without shell crossing, one can adopt the ansatz $r_{f}=\Gamma r_{i}$, with $\Gamma=\mbox{const}$ (Blumenthal, 1986), and Eqs.~\eqref{collapse_halo1} and~\eqref{collapse_halo2}
can be solved numerically for the contraction factor $\Gamma$.
However, in order to be able to mitigate or even switch off the halo adiabatic contraction, we modify by hand the relation between
$r_i$ and $r_f$ and assume, as in Dutton et al. 2007
\begin{equation}
r_{f}=\Gamma^\mu r_{i}\,,
\end{equation}
where $\mu$ is a free phenomenological parameter. Therefore, $\mu=1$ corresponds to adiabatic contraction as in Blumenthal, 1986, while $\mu=0$ completely switches off adiabatic contraction.

%% file: results.tex
\section{Results}

%Evolution of the scale radius relation...
\begin{figure*}
  \centering
  \includegraphics[type=eps,ext=.eps,read=.eps, height=18cm, angle=90]{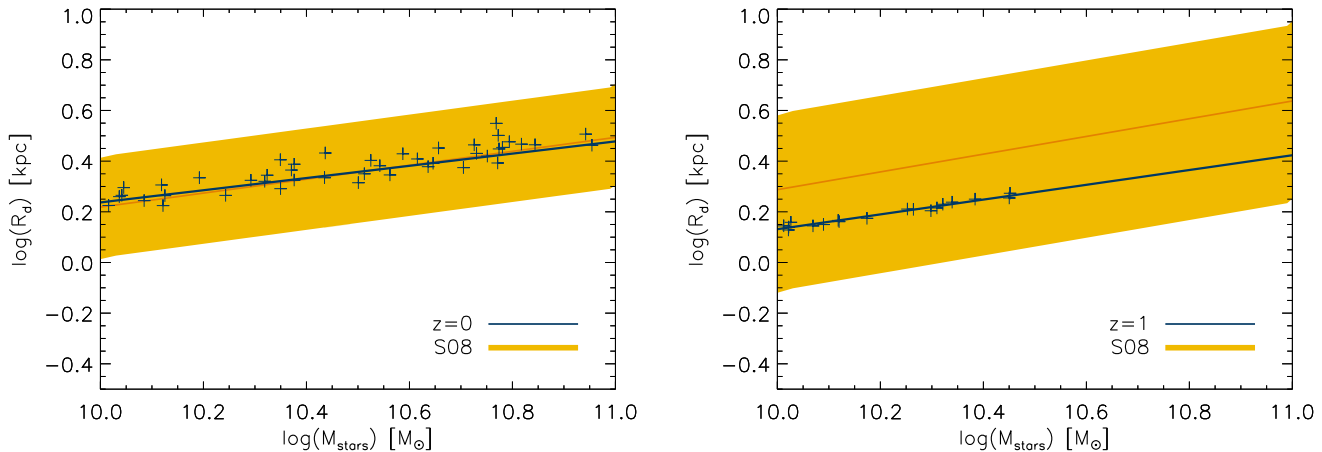}\\
  \caption{The disc size as a function of stellar mass at $z=0$ (left panel) and $z=1$ (right panel) as compared to the results derived in Somerville et al. 2008a and based upon SDSS galaxies (Blanton et al., 2005) locally, and GEMS galaxies (Rix et al., 2004, Caldwell et al., 2005) at high redshifts. Finding good agreements between model and observation at $z=0$ but noting a slight offset when compared to the $z=1$ sample. Yellow shaded regions represent the scatter in disc scale lengths at fixed stellar mass, containing the 95th percentiles. Orange lines show the mean values for the size-mass relation.}
\end{figure*}

In order to make comparisons with observations, we produce a statistical sample of approximately $1000$ galaxies with $z=0$ virial masses in logarithmic increments in the range $9.5<log(M_{v})<13.5$. In order to account for the cosmological abundances of galaxies, we assign each DM halo a weight using the galaxy halo mass function ($\rm GHMF$). This was originally derived by Shankar et al. 2006 in order to account for the one-to-one relationship between galaxies and their host DM haloes. Essentially it is derived using numerically and constrained DM halo mass functions (see Sheth \& Tormen, 2002 ($\rm STMF$), Jenkins et al. 2001) but accounting for the halo occupation distribution (HOD) within DM haloes, which is unity for the majority of galaxy hosting systems, but rapidly increases in haloes within haloes with $M_{v}>10^{13}M_{\odot}$. Thus, to account for this, we subtract the subhalo mass function ($\rm SHMF$) as derived in van den Bosch et al. 2005. The benefit of using these parameterisations is that they are also defined at $z>0$ and thus may be used to extract the number densities of galaxies at higher redshifts, within this work, we use the following:

\begin{equation}\label{1}
    \rm{SHMF}(\psi,z) = \frac{\gamma}{\beta \Gamma(1-\alpha)} \left( \frac{\psi}{\beta} \right)^{-\alpha} \exp\left(-\frac{\psi}{\beta}\right)
\end{equation}

Where $\alpha$, $\beta$ and $\gamma$ are given in van den Bosch et al. 2005, and $\psi=m/M$ are the normalised subhalo masses. Thus $\rm{GHMF}(M,z) = \rm{STMF}(M,z)-\rm{SHMF}(M,z)$ is essentially identical to the $\rm STMF$ at masses below $10^{13}$ at $z=0$, with an exponential cutoff at higher masses due to the dominance of groups and clusters of galaxies. In Fig.1, we show our derived $\rm GHMF$ at different redshifts, as can be seen, at $z=0$ the $\rm GHMF$ is identical to the $\rm SFMF$, but at $M_{v}>10^{12}M_{\odot}$ the $\rm GHMF$ exponentially drops, having a negligible probability at $M_{v}>10^{13}M_{\odot}$. Since clusters form at relatively late times within the standard hierarchical picture, we do not see significant evolution in this cutoff mass to high redshifts, however we do see the typical evolution in the mass function.

\subsection{Galaxy stellar mass function Evolution}

One of the fundamental constraints on the physical mechanisms governing the evolution of luminous matter in galaxies is encoded within the stellar mass function, since its shape holds an imprint of the underlying physics which dominates on different mass scales. Typically, the mass function is fit accurately by a 'Schechter' function (Schechter, 1976) with a low mass power law slope $\alpha$, a characteristic mass $M_{\star}$, and normalisation $\Phi_{\star}$.

It is generally understood that the low mass power law slope may be matched with a combination of ionizing UV background suppression of infall and supernovae feedback, since the potential wells of their host DM haloes are relatively shallow and cannot capture and retain baryonic material (see Benson et al. 2002). Whereas the bright end has proved to be more of a challenge, and is now understood as a combination of cooling inefficiencies coupled with multiple occupation and strong energetic feedback from a central SMBH (Granato et al. 2004, Bower et al. 2006, Croton et al. 2006). These theoretical predictions, however, have been shown to show some discrepancies at higher redshifts (see Marchesini et al. 2008, Fontanot et al. 2009, Kitzbichler \& White, 2007, De Lucia \& Blaizot, 2007)

From an observational perspective Cole et al. 2001, and Bell et al. 2003a used near-IR colours in order to determine the stellar masses, however, more recent approaches model masses using multi-wavelength approaches (Drory et al. 2004, 2005, Fontana et al. 2006, Perez-Gonzalez et al. 2008, Marchescini et al. 2008), exploiting broad-band photometry to compare with libraries of synthetic spectral energy distributions (SEDs) which output the best fitting photometric redshift, stellar mass and SFR. Thus, the determinations of stellar masses is subject to several model-dependent uncertainties and simplifications (such as a smooth star formation history interspersed stochastically with starburst events, unlike theoretical models, which typically exhibit complex histories) and the results are therefore subject to several potential biases (see Marchesini et al. 2008 for an extensive analysis).

Within this work, we compare model predictions between $0<z<4$ with the results of Cole et al. 2001 for local galaxies, Marchesini et al. 2008 for ($z>0$) populations as shown in Fig.2. Locally we find a good agreement with the observed mass function in both high mass cutoff and normalisation, consistently reproducing the observations down to $M_{stars} \approx 10^{10} M_{\odot}$. However, in the lowest mass systems, we do find a slight discrepancy, overproducing the number of low-mass galaxies (an effect which may also be seen in other SAMs, see Fontanot et al. 2009 Fig.1). At higher redshifts we are able to generate a close match to the high mass cutoff up to $z \approx 4$, we view this as a notable success of our model since several other current SAMs find this difficult (see Marchesini et al. 2009, Fig.13), typically under-predicting the cutoff mass and overproducing the number of low mass galaxies. We do, however, find that at $z>0$ our model generates too many low mass galaxies which manifests clearly in the lowest mass systems, however, these mass scales are beyond the range of the observational constraints and thus it remains unclear as to the true faint-end slope at higher redshifts.

Qualitatively, we may view the successful reproduction of the high-mass cutoff as a manifestation of the direct formation of spheroid-SMBH systems at high redshifts. Very early collapse onto spheroid structures without prior disc-formation results in the growth of large galaxies at early times, allowing for the high-mass end of the mass function to be in-place already at $z>4$, in broad agreement also with the concept of 'cosmic downsizing', however, we do find that the overall Schecter function fit poorly describes the model and observation at $z=3.5$, and therefore a comprehensive analysis of the exponential cutoff cannot yet be achieved.

\begin{figure*}
\begin{center}
  \includegraphics[type=eps,ext=.eps,read=.eps, height=18.0cm, angle=90]{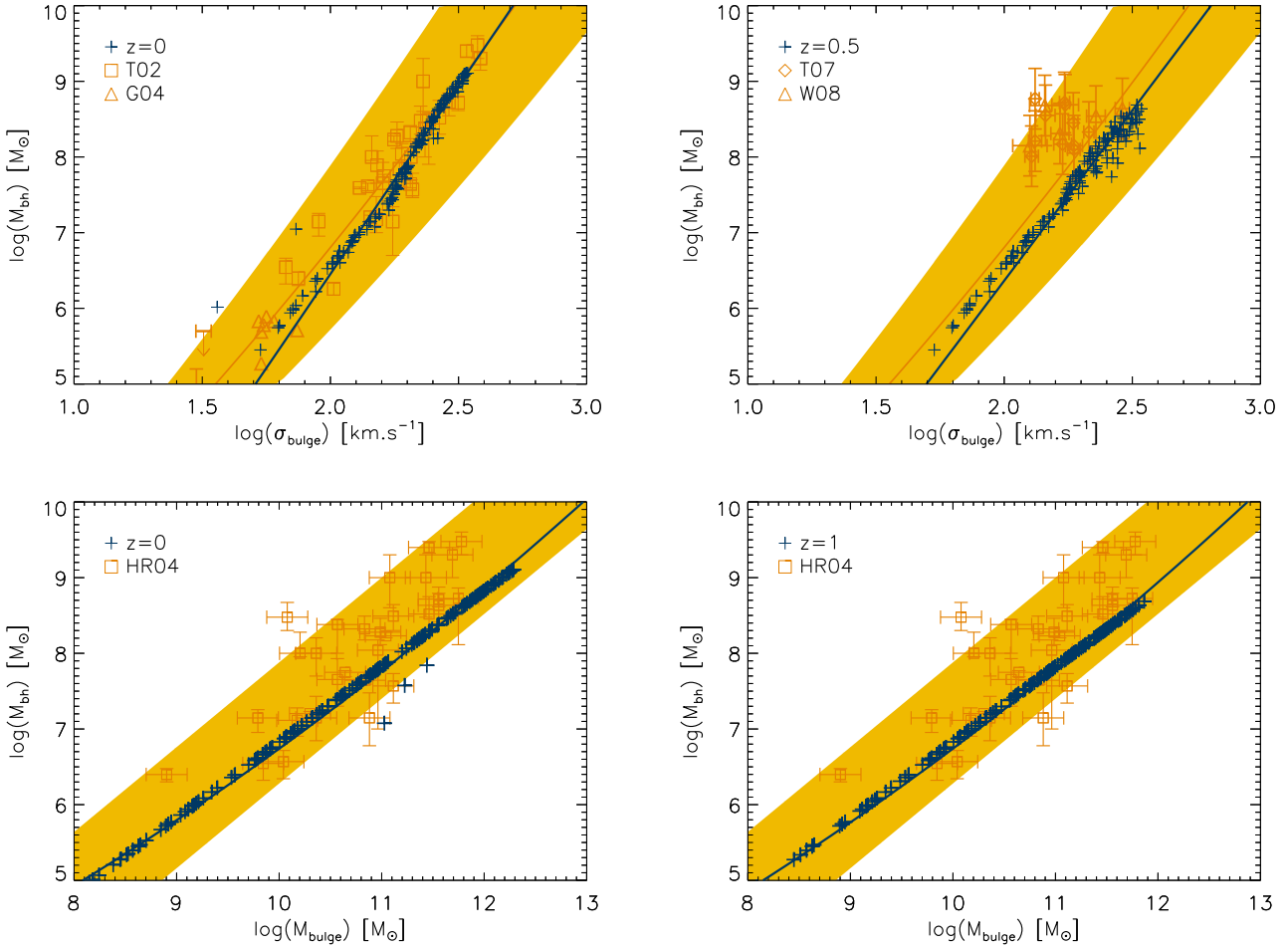}\\
  \caption{The spheroid black-hole scaling relations locally and at $z>0$. The local M-sigma relation as output by model and compared to the results of Tremaine et al. 2002, Greene et al. 2004 (top left) shows agreement with observations over the entire constrained range, and at $z=0.5$ (top right) we compare model predictions with Treu et al. 2007, Woo et al. 2008, finding no significant evolution. For the mass relationship between bulge and SMBH, we compare at $z=0$ to the observations of Haring \& Rix, 2004, finding promising agreements (bottom left), and we predict the form of this relation at $z=1$, also comparing to the $z=0$ sample, and finding no significant evolution (bottom right). Yellow shaded regions represent the errors in the functional fits to observations (orange lines) in all panels.}
\end{center}
\end{figure*}

\begin{figure*}
  \centering
  \includegraphics[type=eps,ext=.eps,read=.eps, height=18cm, angle=90]{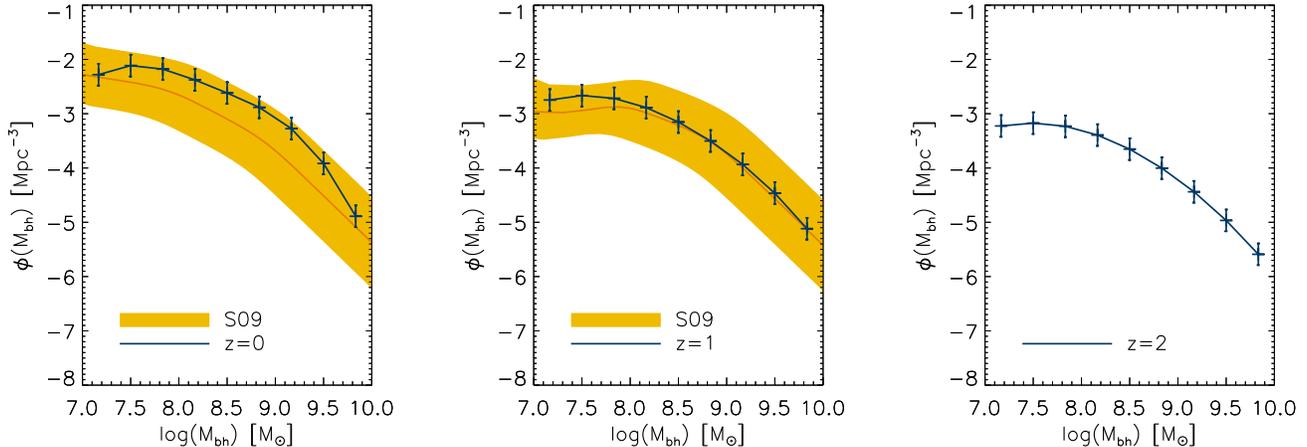}\\
  \caption{The evolution of the black hole mass function as compared to the determinations of Shankar et al. 2009. At $z=0$ we find encouraging agreements (left panel) across the observational range, at $z=1$ we find that we slightly under-predict the number of high mass SMBHs (center panel) but note that this discrepancy may be related to the critical assumption of the validity of the Magorrian (1998) relation at $z>0$. Finally we predict the mass function at $z=2$ (right panel), finding a further drop in normalisation and a lower cutoff. Yellow shaded regions represent the errors in the functional fits to observations (orange lines) and model error bars represent the statistical uncertainties in mean averages, see Table A1. for tabulated data.}
\end{figure*}

\subsection{Massive galaxy number count evolution}

In order to further quantify the growth and evolution of the largest galaxies in the Universe, we compare model predictions with the massive galaxy number density evolution observations between $0<z<5$ by Drory et al. 2005, who found that the number density of the most massive systems evolves in a manner similar to he evolution of lower mass systems and are present at all redshifts within their range. They derive this result by obtaining the stellar mass function for a sample of multicolour observations in the FORS Deep Field (Heidt et al. 2003) and the GOODS-south survey (Giavalisco et al. 2004). By fitting Schechter-functions to the observations at a number of redshift intervals and integrating the results they determine the total stellar mass density evolution, and the galaxy number count evolution. Despite the significant uncertainties due to model-dependent stellar mass determinations, and Schecter-fitting, a striking relation has been obtained, showing that the largest mass systems are being formed at all redshifts within their range, and at $z=5$ a significant number of large mass systems are already formed.

Outputting model predictions, we see in Fig.3 that we reliably reproduce the number densities of the largest ($M>10^{11}M_{\odot}$) systems at all redshifts, whereas we make slight under-predictions of the numbers of intermediate mass systems. Again, we attribute this success to our relaxation of the 'dissipationless collapse' scenario, whereby disc formation and mild star formation are assumed to occur upon gaseous collapse at all epochs. However, we do also note that our under-prediction of the number density in the lower mass ranges is a cause for further analysis.

\subsection{Cosmological star formation rate density}

%revised part here

The cosmological star formation rate density ($\rho_{sfr}$) evolution of the Universe (i.e., the global rate of star formation as a function of redshift) is a key constraint for theoretical models of galaxy formation and cosmology, indicating a clear evolutionary link between the star forming properties of galaxy populations over different redshifts.

The 'Madau diagram', (Madau et al. 1996) has been used as a tool for constraining galaxy evolution models, however it's determination observationally is far from straightforward due to large systematic errors in extracting the SFR from luminosities, and correcting for dust obscuration and incompleteness. These factors led early determinations of the diagram to show a rapid increase by approximately an order of magnitude in $\rho_{sfr}$ from $0<z<1.5$ followed by a peak at $z\approx1.5$ then a steady decline at higher redshifts (see Madau et al. 1996 Fig.9). However, more sophisticated dust modeling and more complete samples at $z>1.5$ resulted in revised estimates of the high-redshift decline; showing a relatively flat $\rho_{sfr}$ out to high-redshifts (Steidel et al. 1999). Combined with observations in the far-IR and sub-mm at intermediate and high redshifts (Hughes et al. 1998, Flores et al. 1999) the redshift dependence of the SFR density has become relatively well constrained to $z \approx 5$.

Originally theoretical attempts to predict the 'Madau diagram' were unable to model the correct evolution (see Cole et al. 1994), however, later works were able to match the results to a good accuracy (including the high-redshift decline, see Cole et al. 2000), and even within the latest generation of SAMs, a modest decline is observed between $2>z>5$ (Somerville et al. 2008b), unlike the most recent observational constraints showing a near-flat evolution to $z\approx6$ (see Hopkins, 2004).

Within this work we utilize the observational compilation in Somerville et al. 2001, which discusses all the aforementioned systematics and corrects for them accordingly (see references therein), also this work accounts for the standard cosmology. Shown in Fig.4, by fitting a cubic polynomial through the data shows (with considerable scatter), a general behavior of a rise in $\rho_{sfr}$ from $0<z<2$ followed by a flattening at $2<z<3$ and a slow decrease to higher redshifts, dropping to the $z=0$ value at $z \approx 5.5$. Outputting the total model $\rho_{sfr}$, we see an overall agreement within the observational range, matching all the observational features. We physically interpret the increase in $\rho_{sfr}$ between $0<z<2$ to several factors, increasingly rapid growth of DM haloes at higher redshifts allows more infalling material and at $z \approx 2$ over the mass range of galactic haloes we have the synchronous formation of spheroid and disc components (since approximately half of the haloes within our sample are in the 'fast collapse' phase and vice versa). Overall, above $z\approx2$ we are dominated by the growth of spheroid-SMBH systems through the dissipative condensation of gas within DM haloes, therefore we typically have higher $\rho_{sfr}$ than predicted by SAMs constructed upon the disc-merger scenario. This naturally gives rise to a slow decrease in  $\rho_{sfr}$ to high redshifts.

Also, we plot $\rho_{sfr}$ separated into both the bulge and disc components, since it has been suggested that they typical 'Hubble-type' morphological classification may be better understood as resulting from differing superpositions of spheroid and disc components (see Driver et al. 2006) where spheroids and discs form two separate classes each with their own distinct formation epochs and mechanisms. We find, in broad agreement with archeological studies, that $\rho_{sfr}$ is dominated by the spheroid component until $z\approx1$, and becomes progressively dominated by the disc component at $z<1$, in accordance with the view that spheroids (and galaxy bulges) are typically 'red and dead', with old stellar populations, whereas discs show ongoing star formation over longer durations.

\subsection{Evolution of the galaxy scaling relations}

As a step beyond simply predicting the accumulation of stellar matter within galaxies, theoretical models are able to make predictions about the dynamics and structure of galaxies which form, allowing for a further level of predictions and constraints from observations. Initial observational advances in this direction came from studies of observable properties of galaxies. Tully \& Fisher, 1977, showed that a tight correlation existed between galaxy luminosity and maximum rotational velocity, these determinations have been confirmed and constrained in a number of latter works (see Haynes et al. 1999).

The Tully-Fisher relation (TFR) thus provides a link between luminous matter (stellar mass) and dynamical matter (total gravitational mass) of galaxies, providing strong constraints on the link between the underlying DM potential and the baryonic matter. Unfortunately however, theoretical attempts to interpret this relation within the framework of full SAMs have found many difficulties; offsets to within 30\% are generally predicted by models (Cole et al. 2000) reinforcing the fact that simultaneous predictions of the stellar mass budgets and the TFR provide tight constraints on models. This is further complicated since 'typical' SAMs make several assumptions and approximations in order to predict the maximum rotational velocity (see Cole et al. 2000). Within our model we directly compute the rotation curve for the composite system given the density distributions of the DM halo, disc, bulge and central reservoir-SMBH system, providing us with a detailed output. Following this, in accordance to observational methods, we output the value for the total rotation curve at 2.2 scale radii, typically corresponding to the 'peak' value for the rotation. Using this value, and plotting against the total galaxy mass in Fig.5 we output the TFR's at three different redshifts.

Comparing model results at $z=0$ to Bell \& de Jong, 2001, and correcting for the stellar mass determinations due to different IMF choices\footnote{Within Bell \& de Jong, 2001, they take values for mass to light ratio which are approximately $30\%$ lower than the Salpeter value, which we account for when comparing stellar masses within this work.}, we are able to make a good match within the observational range (see also C09a), this indicates clearly that the model prescriptions which govern the baryon-to-DM ratios, and the structure of DM haloes are producing the correct dynamical properties at fixed stellar mass. We note, however, that we have investigated the effects of DM halo contraction due to the condensation of baryonic material, however, we find that we most accurately fit to observational results without this effect (setting $\Gamma=0$ see Eqn.13). This interesting finding has also been confirmed in several works focusing on the detailed structural properties of galaxies (see Dutton et al. 2007,2008), concluding that either a low M/L, or no halo contraction may be the only viable routes to achieving simultaneous fits to the TFR and mass functions. Noting also the preliminary work presented in Mo \& Mao, 2004, whereby haloes may become 'pre-processed' due to an early rapid infall of matter, and ensuing mass outflow (through feedback) is able to considerably reduce halo concentrations (see \S2.2 of Mo \& Mao, 2004, see also El-Zant, Shlosman \& Hoffman, 2001,2004, Tonini et al. 2006). We hope to further quantify these effects in a subsequent work.

Outputting model results at higher redshifts we find little evolution between $0<z<1$, and, comparing results to the observational determinations of Conselice et al. 2005 we find a good agreement, indicating that a general 'inside out' growth of discs, coupled with a growing DM halo results in galaxies that typically evolve along the TFR, thus showing little evolution. Despite the small observational sample size (18 galaxies), we also show the $z=2$ TFR as output by our model and compared to Cresci et al. 2009. Within their work, using the SINS survey (Forster Schreiber et al. 2006) which uses integral field spectroscopy to measure the dynamics of high-z galaxies, finding large rotating systems already in place at $z>2$. We find that overall our model shows discrepancies with this data, however, we also note that during these epochs the standard morphological sequence of galaxies is yet to be formed, and many galaxies within this sample are not in an equilibrium state but merely 'rotationally dominated'.

Additionally to the TFR, we show in Fig.6 the evolution of the disc size relation. Basing on the work of Somerville et al. 2008a, who, motivated by detailed observations showing no significant evolution in the relationship between radial size and stellar mass from $z\approx1$ to the present day, conducted a study of theoretical model predictions. By comparing our model predictions to results compiled from the Sloan Digital Sky Survey (SDSS), (Blanton et al. 2005, Somerville et al., 2008a) at $z=0$ we find a strong relation between the disc scale radius (defined to be $R_{d}=0.5959R_{e}$\footnote{We note that there are sevearl cautionary details when converting half-light radii into stellar exponential scale lengths. Firstly, using sersic fitting methods may result in erroneous results for non-exponential discs, and secondly, fitting using circular models generates size biases for inclined galaxies. See Blanton et al., 2005 for more details.}, see Courteau et al. 2007) and the disc stellar mass over the observational range. At $z=1$ however, we find that we systematically under predict the scale radius with an offset of $\approx1.5$ compared to observations of GEMS galaxies (Rix et al., 2004, Caldwell et al., 2005, Somerville et al., 2008a), generating galaxies which are slightly too small at fixed stellar mass compared with observations. These subtle effects, however, are interesting since they appear to be showing that a scenario whereby an initial baryonic collapse and accompanying outflows thus lowering DM halo concentration may help to alleviate these issues (see Mo \& Mao, 2004).

\begin{figure}
\centering
  \includegraphics[type=eps,ext=.eps,read=.eps, height=8.5cm, angle=90]{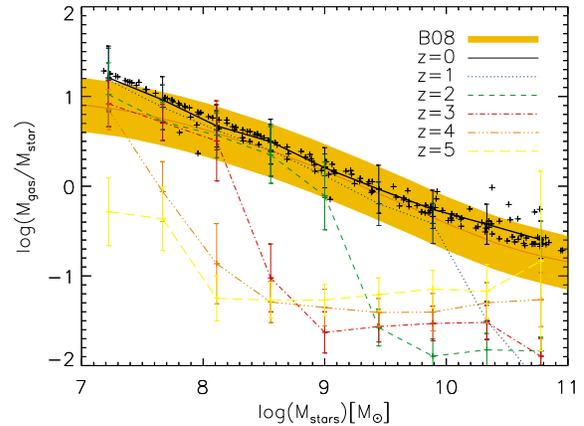}\\
  \caption{The evolution of the gas-star fraction at different stellar masses, comparing model results to Baldry et al. 2008. By directly outputting the stellar and \hi masses we are able to directly compare model results, finding good agreement at $z=0$, we also show the evolution of this relation to $z=5$ showing that low mass galaxies move onto the $z=0$ relation at early times, with the larger systems becoming more gas-rich at later times. The shaded yellow region represents the errors in the functional fit to observations (orange lines) and error bars on model outputs representing Poisson uncertainties in the mean values, see Table A2. for tabulated data.}
\end{figure}

\subsection{Evolution of the black hole scaling relations}

Since the coevolution of a central SMBH and galaxy is a central importance mechanism in the formation and evolution of our system, we show here the evolution of the black hole scaling relations. In the local Universe it is now well established that most galactic nuclei host a central SMBH (Kormendy \& Richstone, 1995). Each SMBH correlates strongly in mass with the global properties of the spheroid component of the host galaxy (Magorrian et al. 1998, Ferrarese \& Merritt, 2000), therefore a theoretical understanding of these relationships is of fundamental importance to galaxy formation theories, and has been shown to provide the key in order to account for the suppression of the formation of massive galaxies (see Granato et al. 2004, Croton et al. 2006, Bower et al. 2006, Ciotti \& Ostriker, 2007). In Fig.7 we compare model predictions of SMBH masses with both the velocity dispersion ($\sigma_{bulge}$)\footnote{We compute the velocity dispersion as $\sigma_{bulge} = 0.65V_{bulge}$} and masses of the spheroid components. As can be seen, when comparing to the $z=0$ properties we find a close agreement to the observational constraints, both to the M-$\sigma$ relation (Tremaine et al. 2002) and the mass scaling (Haring \& Rix, 2004). When comparing model predictions to higher redshifts, we utilize the constraints by Treu et al. 2007 \& Woo et al. 2008 who used high resolution imaging in order to determine the SMBH masses and velocity dispersions at $z=0.36$ and $z=0.57$ respectively. Comparing these to model predictions we find a small offset, since we find that we do not have any significant evolution in the relations. However, in order to further constrain models, and determine whether the observed minor offset is physical or an artefact of increased scatter should allow us to further refine our computations (see Woo et al., 2008). Finally, for completeness, we show our prediction for the mass scalings at $z=1$, again showing that there is no evolution in our models.

Physically, this is due to the rapid growth of SMBH's within galaxies which fix the scaling relations, followed by periods of dormancy. We therefore expect scatter to increase at higher redshifts due to observations of galaxies which are still in the process of fixing their scaling relations.

\subsection{Black hole mass function evolution}

%convolution with scatter...
Utilising and exploiting the strong relationship between the central SMBH mass and the spheroid mass, along with methods in order to convert quasar (QSO) number counts into accreted mass densities onto central SMBH's, attempts to constrain a SMBH mass function were initially conducted by Small \& Blandford, 1992. Following this, several works ( see Salucci et al. 1999) related the luminosity functions of local AGN's and of galaxy spheroids, in order to accurately determine the local mass function of SMBH's. Several further works using various techniques also made estimations (Aller \& Richstone, 2002, Yu \& Tremaine, 2002, McLure \& Dunlop, 2004, Marconi et al., 2004), however, several contrasting results were produced. Shankar et al. 2004 later developed a robust method to determine the local SMBH mass function. In order to make estimations of the SMBH MF at $z>0$ we utilise the methods outlined in Shankar, Bernardi \& Haiman, 2009. In their work, they generate a SMBH mass function at different redshifts by mapping the stellar mass function at $z>0$ onto a SMBH mass function through a Jacobian transformation, assuming the Magorrian (1998) relation remains valid at higher redshifts.

In order to account for scatter, we convolve model outputs with a Gaussian scatter (0.3dex) and present the results in Fig.8. As can be seen we match the $z=0$ mass function over $\approx 4$ orders of magnitude in mass, however at $z=1$ we see model predictions fall slightly below the number density in the largest systems ($M_{bh}>10^{9}M_{\odot}$. Finally, we show our prediction for the SMBH mass function at $z=2$. Assuming the Magorrian relation to be consistent with the local values is uncertain in these early epochs, and we advertise this as a direct prediction of the model.

\begin{figure}
\centering
  \includegraphics[type=eps,ext=.eps,read=.eps, height=8.5cm, angle=90]{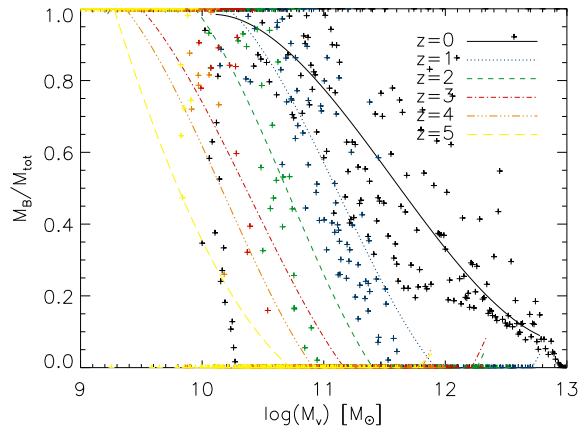}\\
  \caption{The evolution of the galaxy morphologies as a function of virial mass output by our model in the range $0<z<5$. Showing that at any epoch we have galaxies of all morphologies, but the characteristic transition mass decreases with increasing redshift.}
\end{figure}

\subsection{Stellar-to-Gas fraction evolution}

Outputting and analysing the evolution of cold gas within our model, we may make firm predictions to the total neutral gas fraction of galaxies as a function of stellar mass. 
As is shown in Fig.9, comparison to the compiled observations of Baldry, Glazebrook \& Driver, 2008, show an agreement to observations over the entire mass range (4 magnitudes), this encouraging result highlights the accuracy in modeling the conversion of gaseous matter to stellar matter (see C09b for a detailed discussion). In order to make several predictions of our model, we show the evolution of this relation at higher redshifts, finding that globally the gas fraction within galaxies \emph{increases} at lower redshifts, with progressively lower mass galaxies being gas rich at higher redshifts. This is at variance with the general view, not yet observationally verified, that gas fractions increase with higher redshift. For instance, Stewart et al. 2009 assume such a positive z evolution for $M_{gas}/M_{stars}$, based the UV selected sample at z=2 by Erb et al 2006, and their estimate of the gas surface density, based on the Schimdt law. However this determination is indirect and biased toward star forming gas rich systems. We obtain different results from our model partly because in the prediction we include galaxies of all morphologies and evolutionary phase. In our model, an initial collapse causes a large influx of gas into haloes at high redshift, causing the formation of a spheroid-SMBH system within short timescales, with an effectiveness which increases with halo mass. Then high mass systems are soon stripped of gas due to QSO activity, leaving them dormant until a possible secondary disc growth at late times, whilst lower mass systems remain gas-rich because QSO feedback is incapable of removing gas. This results in high-z and high mass gas poor systems which are relatively dormant until conditions become sufficient to support the growth of a disc. As discs grow through the accretion of gas and relatively low star formation rates, the galaxies become progressively more gas-rich, resulting in the tight relation as observed locally. Future more direct and less biased determinations of gas fractions at $z>0$ will represent an interesting test for our model, whose predictions are tabulated in the Appendix.

\subsection{Morphological evolution}

Morphological classification of galaxies has been used as a powerful tool in order to separate galaxies into evolutionary categories. Since Hubble (1926, 1936) defined the classic 'tuning fork' diagram, little modification was required to achieve the modern scheme (see Sandage, 1961). It has become clear that morphological type defines more than merely the appearance of a galaxy, but also highlights general properties of the formation and evolution mechanisms which shape the final galaxy properties, it is understood that this is due to the fundamental underlying galaxy disc and spheroid components which superposed may generate the majority of morphological types (see Driver et al. 2006 for a detailed discussion). Conveniently, galaxy formation models typically separate galaxies into spheroid and disc components and then attempt to translate these into 'early' and 'late' type classifications through post-processed parameterisations, however, these are relatively subjective and may be somewhat arbitrarily chosen in order to 'filter' synthetic galaxy populations (see Cole et al. 2000).

Still little is known about the evolution of galaxy morphology (see Parry, Eke \& Frenk, 2009), however, it is clear that under a physically motivated galaxy formation scenario, whereby spheroids typically comprise of an old, single stellar population with little gas, and discs are a continuously evolving stellar population with a gas rich ISM, the bulge-disc ratio should indicate clearly with several morphological classifications of galaxies. With this in mind, we show in Fig.10 the evolution of the bulge to total mass ratio output by our model, finding that, at $z=0$ we typically produce the observed relationship with low mass DM haloes hosting late-type, disc dominated galaxies, and high mass DM haloes hosting early-type, spheroid dominated galaxies, with a transition close to $L_{\star}$, corresponding to $\approx 10^{11}-10^{12}M_{\odot}$. We find that, at all redshifts all morphologies are present, however, the transition between spheroid dominated galaxies drops to progressively lower masses. We hope to further investigate the morphological evolution, aiming to compare to bulge-disc decomposition studies at higher redshifts as future observations become available.

\subsection{Galactic Archeology}

Plotting the average stellar age of each galaxy against the $z=0$ mass of the galaxy and comparing to the results of Gallazzi et al. 2005.

Finally we compare the average stellar ages of galaxies of different mass, the so called 'galactic archeology'. Observationally Gallazzi et al. 2005 used high resolution SDSS spectra in order to derive estimates for the ages and metallicities of $\approx 170,000$ galaxies through spectral and index fitting to a library of synthetic SEDs. These observations are shown in Fig.11, the large scatter being attributed to the model-dependent age estimation of the stellar populations. As can clearly be seen, the mean stellar age of galaxies decreases with decreasing mass, in contradiction with the naive 'bottom-up' formation scenario, whereby we expect larger galaxies to form at later times. This 'archeological downsizing' has been discussed in detail in Fontanot et al. 2009 (Fig.9), and we find, as with other current SAMs (Wang et al. 2008, Somerville et al. 2008b, Monaco et al. 2007) that we are able to effectively predict the ages of  the largest mass galaxies, but the low mass galaxies form and evolve too early, showing no clear signs of an archeological downsizing. We therefore conclude that our model has difficulties in predicting correctly the properties of the lowest mass galaxies, which typically form too early and are thus contain stellar populations which are too old. We view this as a significant limitation to our model (and to the aforementioned models) and this deserves further analysis.

\begin{figure}
  \centering
  \includegraphics[type=eps,ext=.eps,read=.eps, height=8.5cm, angle=90]{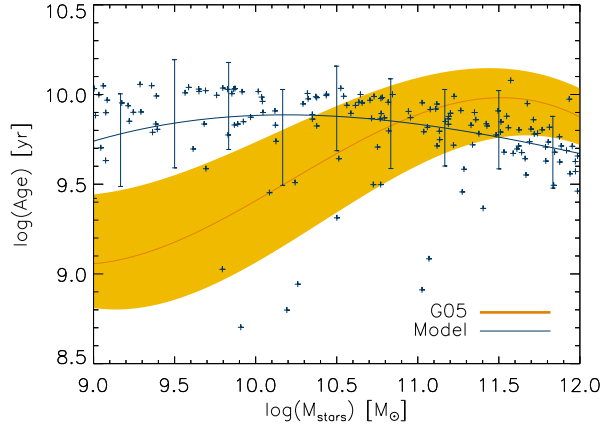}\\
  \caption{The average age of the stellar masses of galaxies as a function of their $z=0$ stellar mass compared to the observations of Gallazzi et al. 2005. We find, as with several other SAMs, that we are able to match the archeological ages of the high mass galaxies, but low mass galaxies clearly form too early in our models, resulting in a significant over-prediction of the ages of these systems. e. The yellow shaded region represents the errors in the functional fit to observations (orange lines) and error bars in the model outputs represent the Poisson uncertainties in the mean averaged values.}
\end{figure}

%% file: conclusions.tex
\section{Conclusions}

Significant recent observational advances have allowed for unprecedented new constraints on the galaxy population, both locally and increasingly high redshifts. Motivated by the emerging phenomenological picture of galaxy evolution, we have presented a theoretical framework in order to interpret several observational constraints, finding a general agreement with several key results, some of which other SAMs find hard to reproduce. Within this work, we have expanded our two-phase galaxy formation model presented in (C09a, C09b) which constitutes a natural extension of the spheroid-SMBH coevolution model presented in Granato et al. 2001, 2004 (see also Granato et al. 2006, Silva et al. 2005, Lapi et al. 2006, 2008, Mao et al. 2007). This model has been shown to naturally reproduce several key results, such as the properties of local elliptical galaxies, the sub-mm galaxy statistics, deep K-band survey results along with the local SMBH mass function and the statistics of high-z QSO's, the local gas fraction, \hi and \hii mass functions, stellar and baryonic mass functions, local luminosity functions (separated into bulge and disc components), and the local Tully-Fisher relation. We inherit these results within this framework.

The basic framework of the model presented here differs significantly from the typical 'disc-merger' scenario for galaxy formation, challenging the assumption that gas cooling and condensation within DM haloes ubiquitously results in the dissipationless collapse onto a disc structure. By allowing for the direct infall onto a spheroid structure at early times we naturally generate large star-burst activity and thus are able to rapidly grow the largest galaxies at high-z, in agreement with many seemingly troublesome observations. We note that the framework presented here is not incompatible with the 'standard' disc merger scenario, but we strongly recommend the assumption of dissipationless collapse at any epoch to form disc structures should be further investigated within the latest SAMs (also noting that within hydrodynamic simulations angular momentum dissipation is commonly seen, Zavala, Okamoto \& Frenk, 2008, Governato et al., 2007).

In order to include the most state-of-the-art processes thought relevant for galaxy formation, we include the effects of cold accretion flows, ionizing UV background radiation, a two-phase ISM, 'radio mode' nuclear feedback, QSO and supernovae feedback, adiabatic response of the DM halo to baryonic structure formation and disc stability criteria. This allows us to directly compare our results with observations and other current SAMs. We find that, after accounting for the one-to-one relationship between DM halo and galaxy properties due to our simplified modeling, we are able to accurately reproduce the stellar mass function in the redshift range $0<z<4$ finding discrepancies only in the very lowest mass haloes (Fig.2). We also show that we are able to match the evolution of massive galaxies from $z=5$, attributing this to the early rapid growth allowed in our models through dissipative collapse onto a spheroid-SMBH structure (Fig.3), and we show the evolution of the cosmological SFR density (Fig.4), finding overall agreement with observations from $z\approx5$, and also showing that SFR in spheroids dominates at $z>1$, and at $z<1$ the Universe favors quiescent disc growth.

Focusing on the galaxy scaling relations, we show (in Fig.5) how the stellar mass Tully-Fisher relation shows little evolution from $0<z<1$, but then a marked difference at $z=2$, and how the disc size evolution also shows little evolution (Fig.6), this naturally results from our models since we naturally generate an 'inside out' growth of discs, where the baryonic and DM evolve together. Also, since we place the mutual feedback between SFR and SMBH growth into central importance for the evolution of the most massive systems, we show in Fig.7 the evolution of the SMBH scaling relations, finding encouraging agreement at $z=0$ and little evolution to higher redshifts, also we show (Fig.8) the SMBH mass function compared to local estimates, and evaluate this at higher redshifts, tentatively comparing it to empirical fitting at $z=1$ and predicting the evolution to $z=2$. In order to further highlight the evolutionary differences in our model, we show in Fig.9 the evolution of the stellar-gas fraction, showing how the $z=0$ relation is constructed, and predict the growth of morphological types in Fig.10. Finally, motivated by recent determinations of mean stellar ages of galaxies, 'archeological downsizing' has been noted in the literature, in Fig.11 we show the mean stellar ages of galaxies as a function of their $z=0$ stellar mass, finding, as with other SAMs, that the smallest systems form too early in our framework (see Fontanot et al. 2009), we confirm that at present this is a robust challenge to all current models.

In summary, under our proposed framework we are able to simultaneously reproduce the vast majority of key observational constraints on galaxy formation in the range $0<z<5$, concentrating particularly on several plots which are notoriously troublesome for SAMs to reproduce, finding minor discrepancies between model and observation mainly where observational results are not constraining and subject to large potential biases. We therefore regard this as a large success of our model. Coupled with the successes of this framework in previous papers, focusing on the detailed properties of galaxies both locally and at high-z, and further advances in observational constraints (particularly with resolved spectroscopy) we hope to further constrain the detailed processes governing the evolution of baryonic matter within evolving DM haloes. 

We also note that there are several points of tension within our model, manifesting within the lowest mass systems. At least in part, these tensions are likely due to oversimplified star formation recipes. Indeed, as discussed in detail in C09b, by including a two-phase ISM, evoking a SFR related to the molecular gas surface density, and including ionizing UV background suppression we are able to significantly reduce the number of low mass satellite galaxies.

However, the main limitation to our simplified computations is that we neglect the environmental effects (tidal stripping and harassment) due to our single MAH approach. We note that the lowest mass galaxies within our observational range are typically the ones most likely to be embedded within larger structures and thus will be subject to external effects (see Mo et al., 2005), by accounting for these we hope to achieve closer matches to the faint-end slope of the stellar mass functions and this may also help to alleviate model discrepancies with the 'archeological downsizing' problem, by pre-heating material and preventing it infalling on the lowest mass haloes at early times due to embedding within larger structures. We thus hope to explore these effects within a subsequent work, noting again that our prescriptions are not mutually incompatible with current SAMs, but simply require modification to loosen the assumption of ubiquitous dissipationless gaseous collapse and naturally ease several tensions between theory and observation.

We therefore advocate the exploration of the evolution of angular momentum in simulations, and hope to conduct a further analysis into how this may be physically modeled within a self-consistent semi-analytical model, basing on the successes of this simplified approach.